\documentclass[preprint,onecolumn,showkeys,preprintnumbers,amsmath,amssymb]{revtex4}
\usepackage{epsf}
\usepackage{dcolumn}
\usepackage{bm}
\everymath{\displaystyle}

\newcommand{\be}{\begin{equation}}
\newcommand{\ee}{\end{equation}}
\newcommand{\bea}{\begin{eqnarray}}
\newcommand{\eea}{\end{eqnarray}}
\newcommand{\bref}[1]{(\ref{#1})}

\begin{document}
\title{Equation of spin motion for a particle with electric and
magnetic charges and dipole moments}

\author{Alexander J. Silenko} \email{alsilenko@mail.ru}
\affiliation{Bogoliubov Laboratory of Theoretical Physics, Joint Institute for Nuclear Research, Dubna 141980, Russia}
\affiliation{Research Institute for Nuclear Problems, Belarusian State University, Minsk 220030, Belarus}

\date{\today}
\begin{abstract}
The general classical equation of spin motion is rigorously derived for a particle with electric and magnetic charges and dipole moments in electromagnetic fields. The equation describing the spin motion relative to the momentum direction in storage rings is also obtained. The importance of the derivation follows from a possible appearance of magnetic charges and EDMs due to the pseudoscalar \emph{CP}-violating electromagnetic interaction caused by dark matter axions.
\end{abstract}

\maketitle

A study of spin motion is one of the most important methods of a search for new physics. 
Witten \cite{EWitten} has shown that a pseudoscalar \emph{CP}-violating electromagnetic interaction leads to a transformation of electrically charged particles to dyons having nonzero electric and magnetic charges. A similar statement has been made in Ref. \cite{AxionCao} in relation to axion-induced electromagnetic effects. In Refs. \cite{Hill,Hill2016,AxionAlexander} and, with some reservations, in Ref. \cite{AxionWang}, an existence of oscillating electric dipole moments (EDMs) 
caused by axion-photon coupling has been stated. Certainly, the presence of real \cite{EWitten} or effective \cite{AxionCao} magnetic charges changes spin dynamics as compared with the known equation (see Refs. \cite{FukuyamaSilenko,PhysScr} and references therein) for particles with EDMs. The problem of a spin motion of a particle with electric and magnetic charges and dipole moments is still unsolved and is rather important because experiments on a search for dark matter axions in spin interactions are now performed by several collaborations \cite{GrahamPhysRevD2018,XENONaxion,axiondJEDI}. In such experiments, a possible presence of magnetic charges should be taken into account. In the present paper, we rigorously derive a general equation of spin motion for a particle with electric and magnetic charges and dipole moments. Certainly, \emph{effective} magnetic charges \cite{AxionCao} do not need to be quantized. For them, the problem of coexistence of electric and magnetic charges does not appear.

We use standard denotations of Dirac matrices (see, e.g., Ref. \cite{BLP}) and the system of units $\hbar=1,~c=1$. We include $\hbar$ and $c$ explicitly when this inclusion clarifies the problem.

The equation of spin motion in electromagnetic fields for a charged (or uncharged) particle with electric and magnetic dipole moments generalizing the well-known Thomas-Bargmann-Michel-Telegdi equation has the form (see Refs. \cite{FukuyamaSilenko,PhysScr} and references therein)
\begin{equation} \begin{array} {c}
\bm \Omega=-\frac{e}{mc}\left[\left(G+\frac{1}{\gamma}\right){\bm B}-\frac{\gamma G}{\gamma+1}({\bm\beta}\cdot{\bm B}){\bm\beta}-\left(G+\frac{1}{\gamma+1}\right)\bm\beta\times{\bm E}\right.\\
+\left.\frac{\eta}{2}\left({\bm E}-\frac{\gamma}{\gamma+1}(\bm\beta\cdot{\bm E})\bm\beta+\bm\beta\times {\bm B}\right)\right].
\end{array} \label{Nelsonh} \end{equation}
Here $G=(g-2)/2,~g=2mc\mu/(es)$, $\eta=2mcd/(es)$, $s$ is the spin number, $\bm\beta=\bm v/c$, and $\gamma$ is the Lorentz factor. 

In the relativistic spin theory, one introduces the spin four-pseudovector and the momentum four-vector, $a^\mu$ and $p^\mu$, whose definition in the particle rest frame is given by \cite{BLP}
\be
a^\mu=(0, {\bm\zeta}), \qquad p^\mu=(m,{\bm 0}).
\ee
So, in any frame
\be
a^\mu p_\mu=0, ~~a_\mu a^\mu=-\bm{\zeta}^2.
\ee
In a frame moving with the velocity ${\bm v}={\bm p}/\epsilon$ ($\epsilon$ is the total kinetic energy), the four-pseudovector $a^\mu=(a^0,~{\bm a})$ is defined by
\be
a^\mu=(a^0,~{\bm a}),\quad {\bm a}=\bm{\zeta}+\frac{{\bm p}(\bm{\zeta}\cdot {\bm p})}{m(\epsilon+m)},\quad a^0=\frac{{\bm a}\cdot{\bm p}}{\epsilon}=\frac{{\bm p}\cdot\bm{\zeta}}{m},\quad {\bm a}^2=\bm{\zeta}^2+\frac{({\bm p}\cdot \bm{\zeta})^2}{m^2}.
\label{spin}
\ee

The Lorentz force $\bm F$ acting on the electric charge $e$ and the Lorentz-like force $\bm F^*$ acting on the magnetic charge $e^*$ are given by \cite{Rindler1989}
\begin{equation}
\frac{d\bm p}{dt}=\bm F+\bm F^*=e(\bm E+\bm\beta\times\bm B)+e^*(\bm B-\bm\beta\times\bm E).
\label{LLL} \end{equation}
The equation of motion depends on the electric and magnetic charges and reads 
\be
m\frac{d u^{\mu}}{d\tau}=eF^{\mu\nu}u_\nu+e^*\widetilde{F}^{\mu\nu}u_\nu,
\label{Lorentz}
\ee
where the dual tensor of electromagnetic field is defined by $\widetilde{F}^{\mu\nu}=\epsilon^{\mu\nu\rho\sigma}F_{\rho\sigma}/2$ and $a_\mu u^\mu=0$. With the denotations of Ref. \cite{LL2}, $F^{\mu\nu}=(-\bm E,\bm B)$ and $\widetilde{F}^{\mu\nu}=(-\bm B,-\bm E)$.

The relativistic equation of spin motion has the general form
\be
\frac{d a^\mu}{d\tau}=A_1 F^{\mu\nu}a_\nu+A_2 u^\mu F^{\nu\lambda}u_\nu a_\lambda+A_3 \widetilde{F}^{\mu\nu}a_\nu+A_4 u^\mu \widetilde{F}^{\nu\lambda}u_\nu a_\lambda.
\label{BMT}
\ee
The coefficients $A_i~(i=1,2,3,4)$ are determined as follows (cf. Ref. \cite{FukuyamaSilenko}).
In the instantaneously accompanying frame (particle rest frame), Eq. \bref{BMT} reduces to
\be
\frac{d a^i}{dt}=\frac{d\zeta^i}{dt}=A_1 F^{ij}\zeta_j+A_3 \widetilde{F}^{ij}\zeta_j=A_1 (\boldsymbol{\zeta}\times {\bf B})^i-A_3 (\boldsymbol{\zeta}\times{\bf E} )^i.
\label{NRspin}
\ee
In this frame, the equation of spin motion for the spin $s=1/2$ is given by
\be
\frac{d\boldsymbol{\zeta}}{dt}=2\mu {\boldsymbol{\zeta}}\times {\bf B}+2d\boldsymbol{\zeta}\times {\bf E}.
\ee
For an arbitrary spin
\be
\frac{d\boldsymbol{\zeta}}{dt}=\frac\mu s {\boldsymbol{\zeta}}\times {\bf B}+\frac ds\boldsymbol{\zeta}\times {\bf E}.
\ee
Comparing this equation with Eq. (\ref{NRspin}), we obtain
\be
A_1=\frac\mu s, \qquad A_3=-\frac ds.
\ee
Values of $A_2,A_4$ result from the equation of motion. 

We should be rather careful. Interactions of the electric charge and dipole with the electric field and the magnetic charge and dipole with the magnetic field are very similar. However, Eq. (\ref{LLL}) shows that interactions of the electric charge with the magnetic field and the magnetic charge with the electric field have different signs. Therefore, the normal (Dirac) magnetic moment originated from the electric charge has the sign opposite to the sign of the ``normal'' EDM caused by the magnetic charge. As a result, the connection between the EDM and the magnetic charge should have the form $G^* =(g^*-2)/2,~g^*=-2mcd/(e^*s)$.

Multiplying Eq. \bref{BMT} by $u_\mu$ and taking into account that $u_\mu u^\mu=1$ leads to
\be
u_\mu\frac{da^\mu}{d\tau}=\left(\frac\mu s+A_2\right)F^{\mu\nu}u_\mu a_\nu+\left(-\frac ds+A_4\right)\widetilde{F}^{\mu\nu}u_\mu a_\nu.
\ee
Since 
\be
\frac{d(a_\mu u^\mu)}{d\tau}=0,\qquad u_\mu\frac{da^\mu}{d\tau}=-a_\mu\frac{du^\mu}{d\tau}=\frac{1}{m}\left(eF^{\mu\nu}+ e^*\widetilde{F}^{\mu\nu}\right)u_\mu a_\nu,
\ee
we obtain
\be
A_2=\frac{e}{m}-\frac\mu s\equiv -\frac{\mu'}{s},\qquad A_4=\frac{e^*}{m}+\frac ds\equiv \frac{d'}{s}.
\ee
Here $d$, $-e^*s/m$, and $d'$ are the total, ``normal'' (Dirac-like), and ``anomalous'' EDMs. This distribution repeates the distribution for the magnetic moments $\mu,~es/m$, and $\mu'$.

As a result, the equation of spin motion takes the form
\be
\frac{da^\mu}{d\tau}=\frac{\mu}{s} F^{\mu\nu}a_\nu-\frac{\mu'}{s}u^\mu F^{\nu\lambda}u_{\nu}a_\lambda-\frac{d}{s}\widetilde{F}^{\mu\nu}a_\nu+\frac{d'}{s}u^\mu \widetilde{F}^{\nu\lambda}u_\nu a_\lambda.
\label{BMT1}
\ee
Unlike the Thomas-Bargmann-Michel-Telegdi equation \cite{Thomas,BMT}, this generalized equation takes into account the effective \cite{AxionCao} or real \cite{EWitten} magnetic charge and the EDM.

In the above derivation, we followed Ref. \cite{FukuyamaSilenko}. Next calculations utilize the approach by Jackson \cite{Jackson} and almost repeate Ref. \cite{PhysScr}. 

With the use of Eq. (\ref{Lorentz}), the obtained equation can be presented in the form
\begin{equation}
\frac{da^\mu}{d\tau}=\frac{\mu}{s}\left(F^{\mu\nu}a_\nu-u^\mu F^{\nu\lambda}u_{\nu}a_\lambda\right)-\frac{d}{s}\left(\widetilde{F}^{\mu\nu}a_\nu-u^\mu \widetilde{F}^{\nu\lambda}u_\nu a_\lambda\right)-
u^\mu \frac{du^\lambda}{d\tau}a_\lambda.
\label{BMTn} \end{equation}

It is convenient to denote
\begin{equation}
\Phi^\mu=\frac{\mu}{s}\left(F^{\mu\nu}a_\nu-u^\mu F^{\nu\lambda}u_{\nu}a_\lambda\right)-\frac{d}{s}\left(\widetilde{F}^{\mu\nu}a_\nu-u^\mu \widetilde{F}^{\nu\lambda}u_\nu a_\lambda\right).
\label{Jackson1} \end{equation}

Evidently, $\Phi^\mu=(\Phi^0,\bm \Phi)$ is a four-vector. Since $u_\mu\Phi^\mu=\gamma(\Phi^0-\bm\beta\cdot\bm\Phi)=0$, it satisfies the relation $\Phi^0=\bm\beta\cdot\bm\Phi$. The last term in Eq. (\ref{BMTn}) can be transformed as follows \cite{Jackson}:
\begin{equation}
u^\mu \frac{du^\lambda}{d\tau}a_\lambda=-u^\mu \gamma\bm a\cdot\frac{d\bm\beta}{d\tau}.
\label{Jackson2} \end{equation}
Thus, Eq. (\ref{BMTn}) leads to
\begin{equation}
\frac{da^0}{d\tau}=\Phi^0+\gamma^2\bm a\cdot\frac{d\bm\beta}{d\tau}, \qquad \frac{d\bm a}{d\tau}=\bm\Phi+\gamma^2\bm\beta\left(\bm a\cdot\frac{d\bm\beta}{d\tau}\right).
\label{Jackson3} \end{equation}

The equation of motion for the rest-frame spin $\bm\zeta$ can be calculated with the relations
$$ 
{\bm \zeta}=\bm{a}-\frac{\gamma}{\gamma+1}{\bm\beta}({\bm\beta}\cdot\bm{a}),\qquad\frac{d}{d\tau}\left(\frac{\gamma}{\gamma+1}{\bm\beta}\right)=\frac{\gamma}{\gamma+1}\frac{d\bm\beta}{d\tau}+
\frac{\gamma^3}{(\gamma+1)^2}{\bm\beta}\left({\bm\beta}\cdot\frac{d\bm\beta}{d\tau}\right).
$$ 
Therefore, the equation of spin motion has the form (cf. Refs. \cite{PhysScr,Jackson})
\begin{equation}
\frac{d\bm\zeta}{d\tau}=\bm\Phi-\frac{\gamma\bm\beta}{\gamma+1}\Phi^0+
\frac{\gamma^2}{\gamma+1}\bm\zeta\times\left({\bm\beta}\times\frac{d\bm\beta}{d\tau}\right).
\label{Jackson4} \end{equation}

The transformation of the given four-vector $\Phi^\mu$ to the instantaneously accompanying frame results in
$\bigl(\Phi^{(0)}\bigr)^\mu=\bigl(0,\bm\Phi^{(0)}\bigr)$, where
$$\bm\Phi^{(0)}=\bm\Phi-\frac{\gamma}{\gamma+1}\bm\beta(\bm\beta\cdot\bm\Phi)=\bm\Phi-\frac{\gamma\bm\beta}{\gamma+1}\Phi^0.$$ We should differ $\Phi^0$ and $\bm\Phi^{(0)}$.

Since $dt=\gamma \,d\tau$, the derivation of $\bm\Phi^{(0)}$ from Eq. (\ref{Jackson1}) leads to the following equation of spin motion: 
\begin{equation}
\frac{d\bm\zeta}{dt}=-\left(\frac{d\bm E^{(0)}}{s\gamma}+\frac{\mu\bm B^{(0)}}{s\gamma}\right)\times\bm\zeta-
\frac{\gamma^2}{\gamma+1}\left({\bm\beta}\times\frac{d\bm\beta}{dt}\right)\times\bm\zeta.
\label{Jackson5} \end{equation}
The angular velocity of spin precession is given by \cite{PhysScr,Jackson}
\begin{equation}
\bm\Omega=-\left(\frac{d\bm E^{(0)}}{s\gamma}+\frac{\mu\bm B^{(0)}}{s\gamma}\right)-
\frac{\gamma^2}{\gamma+1}\left({\bm\beta}\times\frac{d\bm\beta}{dt}\right)=\bm\Omega^{(0)}+\bm\omega_T,
\label{angvelo} \end{equation}
where 
\begin{equation}   \begin{array}{c}
\bm\Omega^{(0)}=-\frac{d\bm E^{(0)}}{s\gamma}-\frac{\mu\bm B^{(0)}}{s\gamma}
\end{array} \label{multanv} \end{equation}
and $\bm\omega_T$ is the angular velocity of the Thomas precession:
\begin{equation}\begin{array} {c}
\bm\omega_T=-\frac{\gamma^2}{\gamma+1}\left(\bm \beta\times\frac{d\bm \beta}{dt}\right).
\end{array}\label{Thompre}\end{equation}

Therefore, the total angular velocity of spin precession is the sum of two parts. The first part is given by the Lorentz transformation between the instantaneously accompanying frame and the lab frame. The fields in the instantaneously accompanying frame are equal to \cite{Jackson} 
\begin{equation}\begin{array} {c}
\bm E^{(0)}=\gamma\left[\bm E-\frac{\gamma}{\gamma+1}\bm\beta(\bm\beta\cdot\bm E)
+\bm \beta\times\bm B\right],\\
\bm B^{(0)}=\gamma\left[\bm B-\frac{\gamma}{\gamma+1}\bm\beta(\bm\beta\cdot\bm B)
-\bm \beta\times\bm E\right].
\end{array} \label{meffinal} \end{equation}
The second part is the contribution from the Thomas precession. This part defines the additional spin precession caused by a purely kinematical effect of a rotation of the particle rest frame (see, e.g., Refs. \cite{Jackson,Rindler}). Only the second part depends on the magnetic charge.

The particle acceleration expressed in terms of the lab frame fields is defined by Eq. (\ref{LLL}). The total acceleration reads
\begin{equation}\begin{array} {c}
\frac{d\bm\beta}{dt}=\frac{e}{mc\gamma}\left[\bm E+{\bm\beta}\times {\bm B}-\bm\beta(\bm\beta\cdot{\bm E})\right]+\\
\frac{e^*}{mc\gamma}\left[\bm B-{\bm\beta}\times {\bm E}-\bm\beta(\bm\beta\cdot{\bm B})\right].
\end{array} \label{eqm} \end{equation}
With the use of Eqs. (\ref{meffinal}) and (\ref{eqm}), one can bring Eqs. (\ref{multanv}) and (\ref{Thompre}) to the form
\begin{equation} \begin{array} {c}
\bm \Omega^{(0)}=-\frac{\mu}{s}\left[\bm B-\frac{\gamma}{\gamma+1}{\bm\beta}({\bm\beta}\cdot{\bm B})-\bm\beta\times{\bm E}\right]\\ -\frac{d}{s}\left[\bm E-\frac{\gamma}{\gamma+1}{\bm\beta}({\bm\beta}\cdot{\bm E})+\bm\beta\times{\bm B}\right],\\
\bm\omega_T=\frac{e}{m}\left[\frac{\gamma-1}{\gamma}\bm B-\frac{\gamma}{\gamma+1}{\bm\beta}({\bm\beta}\cdot{\bm B})-\frac{\gamma}{\gamma+1}\bm\beta\times {\bm E}\right]\\ -\frac{e^*}{m}\left[\frac{\gamma-1}{\gamma}\bm E-\frac{\gamma}{\gamma+1}{\bm\beta}({\bm\beta}\cdot{\bm E})+\frac{\gamma}{\gamma+1}\bm\beta\times {\bm B}\right].
\end{array} \label{Nelsonn} \end{equation}
The resulting angular velocity of spin motion reads
\begin{equation} \begin{array} {c}
\bm \Omega=-\frac{e}{m}\left[\left(G+\frac{1}{\gamma}\right){\bm B}-\frac{G\gamma}{\gamma+1}({\bm\beta}\cdot{\bm B}){\bm\beta}-\left(G+\frac{1}{\gamma+1}\right)\bm\beta\times{\bm E}\right]\\
+\frac{e^*}{m}\left[\left(G^*+\frac{1}{\gamma}\right){\bm E}-\frac{G^*\gamma}{\gamma+1}({\bm\beta}\cdot{\bm E}){\bm\beta}+\left(G^*+\frac{1}{\gamma+1}\right)\bm\beta\times{\bm B}\right].
\end{array} \label{Nelsonf} \end{equation}
When $e^*=0$, Eq. (\ref{Nelsonf}) takes the form (\ref{Nelsonh}).
Equation (\ref{Nelsonf}) is useful for experiments with atoms. 

While the magnetic charge and the EDM caused by dark matter axions are oscillating, the angular velocity of their oscillations, $\omega_a=m_ac^2/\hbar$, is very small as compared with $\Omega$ and $\omega$. As a result, it is convenient to determine the spin motion of particles in storage rings relative to the velocity and momentum direction. Such a transformation presents Eq. (\ref{Nelsonf}) in terms of the unit vector in this direction, ${\bm N}={\bm\beta}/\beta=\bm{p}/p$:
$$\frac{d{\bm
N}}{dt}=\frac{\dot{\bm\beta}}{\beta}
-\frac{\bm\beta}{\beta^3}\left({\bm\beta}\cdot\dot{\bm\beta}\right)=\bm\omega\times{\bm N}, \qquad \bm\omega=\frac{e}{m\gamma}\left(\frac{{\bm\beta}\times{\bm E}}{\beta^2}-{\bm B}\right)+\frac{e^*}{m\gamma}\left(\frac{{\bm\beta}\times{\bm B}}{\beta^2}+{\bm E}\right),
$$ where $\bm\omega$ is the angular
velocity of rotation of the velocity and momentum direction. Thus, the angular velocity of the spin rotation relative to this direction (i.e., in the Frenet-Serret coordinate system) is given by
\begin{equation} \begin{array} {c}
\bm\Omega_{FS}=\bm \Omega-\bm\omega=-\frac{e}{m}\left[G{\bm B}-\frac{G\gamma}{\gamma+1}({\bm\beta}\cdot{\bm B}){\bm\beta}-\left(G-\frac{1}{\gamma^2-1}\right){\bm\beta}\times{\bm E}\right]\\+\frac{e^*}{m}\left[G^*{\bm E}-\frac{G^*\gamma}{\gamma+1}({\bm\beta}\cdot{\bm E}){\bm\beta}+\left(G^*-\frac{1}{\gamma^2-1}\right){\bm\beta}\times {\bm B}\right].
\end{array} \label{Nelsonm} \end{equation} The quantities $e^*$ and $d$ (but not $G^*$) oscillate with the frequency $\omega_a$.

For more detailed derivations, especially for calculations of systematical errors, it is helpful to use the cylindrical coordinates \cite{PRStAcBeam}.

In summary, the general classical equation of spin motion has been rigorously derived for particles with electric and magnetic charges and dipole moments in electromagnetic fields. While we do not use any dual transformation, this equation possesses the dual symmetry $e\rightarrow e^*,~G\rightarrow G^*,~\bm E\rightarrow \bm B,~\bm B\rightarrow -\bm E$. The equation is also presented in the form convenient for the description of spin motion in storage rings. The necessity for our derivation follows from Refs. \cite{EWitten,AxionCao} substantiating an appearance of magnetic charges and EDMs due to the pseudoscalar \emph{CP}-violating electromagnetic interaction.

The author is grateful to N.N. Nikolaev for helpful discussions and comments.


\begin{thebibliography}{99}
\bibitem{EWitten}
E. Witten, Dyons of Charge $e \theta/2 \pi$, Phys. Lett. B \textbf{86}, 283 (1979). 

\bibitem{AxionCao}
ChunJun Cao and A. Zhitnitsky, Axion detection via topological Casimir effect, Phys. Rev. D \textbf{96}, 015013 (2017).

\bibitem{Hill}
C. T. Hill, Axion Induced Oscillating Electric Dipole
Moments, Phys. Rev. D \textbf{91}, 111702 (2015).

\bibitem{Hill2016}
C. T. Hill, Axion Induced Oscillating Electric Dipole
Moment of the Electron, Phys. Rev. D \textbf{93}, 025007 (2016).

\bibitem{AxionAlexander}
S. Alexander and R. Sims, Detecting axions via induced electron spin precession, Phys. Rev. D \textbf{98}, 015011 (2018).

\bibitem{AxionWang}
Zihang Wang and Lijing Shao, Axion induced spin effective couplings, Phys. Rev. D \textbf{103}, 116021 (2021).

\bibitem{FukuyamaSilenko}
T. Fukuyama and A. J. Silenko, Derivation of Generalized
Thomas-Bargmann-Michel-Telegdi Equation for a Particle with Electric
Dipole Moment, Int. J. Mod. Phys. A \textbf{28}, 1350147 (2013).

\bibitem{PhysScr}
A. J. Silenko, Spin precession of a particle with an electric dipole moment:
contributions from classical electrodynamics and from the Thomas effect,
Phys. Scripta {\bf 90}, 065303 (2015).

\bibitem{GrahamPhysRevD2018}
P. W. Graham, D. E. Kaplan, J. Mardon, S. Rajendran, W. A. Terrano,
L. Trahms, and T. Wilkason, Spin precession experiments for light axionic dark matter, Phys. Rev. D \textbf{97}, 055006 (2018).

\bibitem{XENONaxion}
E. Aprile \emph{et al.} (XENON Collaboration), Excess electronic recoil events in XENON1T, Phys. Rev. D \textbf{102}, 072004 (2020).

\bibitem{axiondJEDI}
S. Karanth \emph{et al.} (JEDI Collaboration), First search for axionlike particles in a storage ring using a polarized deuteron beam, Phys. Rev. X Phys. Rev. X \textbf{13}, 031004 (2023). 

\bibitem{BLP}
V. B. Berestetskii, E. M. Lifshitz, and L. P. Pitayevskii,
{\em Quantum Electrodynamics}, 2nd ed. (Pergamon, Oxford, 1982).

\bibitem{Rindler1989}
W. Rindler, Relativity and electromagnetism: The force on a magnetic monopole, Am. J. Phys. \textbf{57}, 993 (1989).

\bibitem{LL2}
L. D. Landau and E.M. Lifshitz, \emph{The Classical Theory of Fields},
4th ed. (Butterworth-Heinemann, Oxford, 1980).

\bibitem{Thomas}
L. H. Thomas, The motion of the spinning electron, Nature \textbf{117}, 514 (1926); The kinematics of an electron with an axis,
Phil. Mag. \textbf{3}, 1 (1927).

\bibitem{BMT}
V. Bargmann, L. Michel, and V. L. Telegdi, Precession of the polarization
of particles moving in a homogeneous electromagnetic field, Phys. Rev. Lett. \textbf{2}, 435 (1959).

\bibitem{Jackson}
J. D. Jackson, \emph{Classical Electrodynamics}, 3rd ed. (Wiley, New York, 1998).

\bibitem{Rindler}
W. Rindler, \emph{Relativity: Special, General, and Cosmological}, 2nd ed. (Oxford Univ. Press, Oxford, 2001), pp. 199-200;
R. U. Sexl and H. K. Urbantke, \emph{Relativity, Groups, Particles: Special Relativity and Relativistic Symmetry in Field and Particle Physics}, (Springer, New York, 2001); A. A. Ungar, \emph{Beyond the Einstein Addition Law and its Gyroscopic Thomas Precession}, (Kluwer Acad. Publ., Dordrecht, 2001).

\bibitem{PRStAcBeam} A. J. Silenko, Equation of spin motion in storage rings in the cylindrical coordinate system,
Phys. Rev. ST Accel. Beams \textbf{9}, 
034003 (2006).

\end{thebibliography}
\end{document}